\documentstyle[aps,prl,multicol,epsfig]{revtex}

\begin{document}

\draft

\title{Observation of a Coherence Length Effect in
Exclusive $\rho^0$ Electroproduction}

\author{
\centerline {\it The HERMES Collaboration}
K.~Ackerstaff$^5$,
A.~Airapetian$^{34}$,
N.~Akopov$^{34}$,
I.~Akushevich$^6$,
M.~Amarian$^{34,29,25}$,
E.C.~Aschenauer$^{6,13,14,25}$,
H.~Avakian$^{10}$,
R.~Avakian$^{34}$,
A.~Avetissian$^{34}$,
B.~Bains$^{15}$,
S.~Barrow$^{27}$, 
C.~Baumgarten$^{23}$,
M.~Beckmann$^{12}$,
St.~Belostotski$^{28}$,
J.E.~Belz$^{4,30,31}$,
Th.~Benisch$^8$,
S.~Bernreuther$^8$,
N.~Bianchi$^{10}$,
S.~Blanchard$^{24}$, 
J.~Blouw$^{25}$,
H.~B\"ottcher$^6$,
A.~Borissov$^{6,14}$,
J.~Brack$^4$,
S.~Brauksiepe$^{12}$,
B.~Braun$^{8,23}$,
B.~Bray$^3$,
St.~Brons$^6$,
W.~Br\"uckner$^{14}$,
A.~Br\"ull$^{14}$,
E.E.W.~Bruins$^{20}$,
H.J.~Bulten$^{18,25,33}$,
R.V.~Cadman$^{15}$,
G.P.~Capitani$^{10}$,
P.~Carter$^3$,
P.~Chumney$^{24}$,
E.~Cisbani$^{29}$,
G.R.~Court$^{17}$,
P.~F.~Dalpiaz$^9$,
P.P.J.~Delheij$^{31}$, 
E.~De Sanctis$^{10}$,
D.~De Schepper$^{20}$,
E.~Devitsin$^{22}$,
P.K.A.~de Witt Huberts$^{25}$,
P.~Di Nezza$^{10}$,
M.~D\"uren$^8$,
A.~Dvoredsky$^3$,
G.~Elbakian$^{34}$,
J.~Ely$^4$,
J.~Emerson$^{30,31}$, 
A.~Fantoni$^{10}$,
A.~Fechtchenko$^7$,
M.~Ferstl$^8$,
D.~Fick$^{19}$, 
K.~Fiedler$^8$,
B.W.~Filippone$^3$,
H.~Fischer$^{12}$,
H.T.~Fortune$^{27}$, 
B.~Fox$^4$,
S.~Frabetti$^9$,
J.~Franz$^{12}$,
S.~Frullani$^{29}$,
M.-A.~Funk$^5$,
N.D.~Gagunashvili$^7$,
P.~Galumian$^1$,  
H.~Gao$^{2,15,20}$,
Y.~G\"arber$^6$,
F.~Garibaldi$^{29}$,
G.~Gavrilov$^{28}$,
P.~Geiger$^{14}$,
V.~Gharibyan$^{34}$,
V.~Giordjian$^{10}$,
A.~Golendukhin$^{19,34}$,
G.~Graw$^{23}$,
O.~Grebeniouk$^{28}$,
P.W.~Green$^{1,31}$,
L.G.~Greeniaus$^{1,31}$,
C.~Grosshauser$^8$,
M.G.~Guidal$^{25}$,
A.~Gute$^8$,
V.~Gyurjyan$^{10}$, 
J.P.~Haas$^{24}$, 
W.~Haeberli$^{18}$,
J.-O.~Hansen$^2$,
D.~Hasch$^6$,
O.~H\"ausser\cite{author_note1}$^{30,31}$,
F.H.~Heinsius$^{12}$,
R.S.~Henderson$^{31}$,
Th.~Henkes$^{25}$,
M.~Henoch$^8$,
R.~Hertenberger$^{23}$,
Y.~Holler$^5$,
R.J.~Holt$^{15}$,
W.~Hoprich$^{14}$,
H.~Ihssen$^{5,25}$,
M.~Iodice$^{29}$
% ...
}

\address{
% ...
$^5$DESY, Deutsches Elektronen Synchrotron, 22603 Hamburg, Germany\\
% ...
$^8$Physikalisches Institut, Universit\"at Erlangen-N\"urnberg,
91058 Erlangen, Germany\\
% ...
$^{12}$Fakult\"at f\"ur Physik, Universit\"at Freiburg, 79104 Freiburg,
Germany\\
% ...
$^{14}$Max-Planck-Institut f\"ur Kernphysik, 69029 Heidelberg, Germany\\
$^{15}$Department of Physics, University of Illinois, Urbana,
Illinois 61801, USA\\
% ...
$^{23}$Sektion Physik, Universit\"at M\"unchen, 85748 Garching, Germany\\
% ...
$^{25}$Nationaal Instituut voor Kernfysica en Hoge-Energiefysica (NIKHEF),
1009 DB Amsterdam, The Netherlands\\
% ...
$^{29}$Istituto Nazionale di Fisica Nucleare, Sezione Sanit\`a
and Physics Laboratory, Istituto Superiore di Sanit\'a,
00161 Roma, Italy\\
$^{30}$Department of Physics, Simon Fraser University, Burnaby,
British Columbia V5A 1S6, Canada\\
$^{31}$TRIUMF, Vancouver, British Columbia V6T 2A3, Canada\\
% ...
}

\maketitle

\begin{abstract}
Exclusive incoherent electroproduction
of the $\rho^0(770)$ meson from
$^1$H, $^2$H, $^3$He, and $^{14}$N targets
has been studied by the HERMES experiment
at squared four-momentum transfer
$Q^2>0.4~\rm GeV^2$ and
positron energy loss $\nu$ from 9 to $20~\rm GeV$.
The ratio of the $^{14}$N to $^1$H cross sections per nucleon,
known as the nuclear transparency,
was found to decrease with increasing
coherence length of quark-antiquark fluctuations of
the virtual photon.
The data provide clear evidence
of the interaction of the quark-antiquark fluctuations with
the nuclear medium.
\end{abstract}


\begin{multicols}{2}[]

The space-time evolution
of a virtual quantum state,
such as a quark-antiquark ($q\bar{q}$) fluctuation of a photon,
can be probed by studying its
propagation through a perturbing medium.
The unperturbed virtual state can travel a distance
$l_c$, known as the ``coherence length,''
in the laboratory frame during its lifetime.
The interactions between the state and the medium
can be studied
at different values of $l_c$
by varying the kinematics at which the state is produced.
In this letter, interactions of a $q\bar{q}$ fluctuation with the nuclear
medium are studied by measuring
the nuclear
dependence of the
exclusive $\rho^0$ electroproduction cross section.

Studies of the hadronic ($q\bar{q}$)
structure of high-energy photons started with
ground-work by 
Yang and Mills, Sakurai,
Gell-Mann and Zachariasen,
and Berman and Drell in the early 1960's\cite{namedrop}.
The hadronic structure arises
from fluctuations of the (real or virtual) photon
to short-lived
quark-antiquark states
of mass $M_{q\bar{q}}$ and
propagation distance
$l_c = 2\nu/(Q^2 + M_{q\bar{q}}^2)$\cite{bauer,kopel,gottfried},
where
$-Q^2$ and $\nu$ are the squared mass and laboratory-frame energy of
the photon
(adopting units where $\hbar = c = 1$).
The $q\bar{q}$ fluctuations are assumed to dominate
many photon-induced reactions in the laboratory frame\cite{bauer}.
For example in
exclusive production of the $\rho^0$ meson,
a $q\bar{q}$ pair
is scattered onto
the physical $\rho^0$ mass shell
by a diffractive interaction with the target
\cite{bauer,kopel,gottfried,qcdmodel}.

In nuclear targets,
photon-induced reactions
can be affected by the
initial state
interactions (ISI) of the $q\bar{q}$ states with the nuclear medium.
The ISI are maximized when 
$l_c$ is large
compared to the 
nuclear radius $R_A$, and the photon converts to the $q\bar{q}$ pair
before entering the nucleus
\cite{bauer,kopel,gottfried}.
The hadronic
ISI vanish in the limit $l_c \ll R_A$ of negligible $q\bar{q}$
interaction path.
The dependence of the ISI on $l_c$
can be measured explicitly in
exclusive $\rho^0$ production experiments,
where
a single mass---namely, the $\rho^0$ mass---dominates
$M_{q\bar{q}}$ and $l_c$\cite{bauer,kopel,gottfried}.
Due largely to limited coverage in $l_c$, previous experiments have not yet
seen the expected $l_c$ dependence\cite{bauer,mcclellan}.

In exclusive reactions
a specific final state is produced without
additional particles, for example
$eN \rightarrow e\rho^0N$ (here $N$ is a nucleon).
The effect of the nuclear medium on the particles
in the initial and final states of such reactions
can be characterized
by the nuclear transparency $T_A$.
It is defined as the ratio of the measured cross section to that
expected in the absence of these initial and final state interactions
(ISI and FSI).
If the ISI and FSI amplitudes factorize from the exclusive scattering
amplitude, then
$T_A$
is the probability that no
significant ISI or FSI occur.
The transparency
has been used to study the space-time dynamics of several exclusive
reactions\cite{bauer,mcclellan,oldtrans,e665,nmc}.
This paper reports measurements
of the nuclear transparency for
exclusive incoherent $\rho^0$ electroproduction
on $^2$H, $^3$He, and $^{14}$N targets
at $Q^2 > 0.4~\rm GeV^2$,
$9{~\rm GeV}< \nu < 20~\rm GeV$, and
$0.6{~\rm fm} \lesssim l_c < 8~\rm fm$.
The data provide an explicit demonstration
that the interactions of
the photon with the nuclear medium
depend on the propagation distance $l_c$ of the $q\bar{q}$ pair.

The
data were obtained during the 1995-1997 running periods
of the HERMES experiment
using $^1$H, $^2$H, $^3$He,
and $^{14}$N internal gas targets in
the 27.5~GeV HERA positron storage ring
at DESY.
The scattered $e^+$ and the $\pi^+\pi^-$ pair from the $\rho^0$ decay
($\approx 100\%$
branching ratio)
were detected in the HERMES
forward
spectrometer\cite{hermes}.

The $\rho^0$ production sample was
extracted from events
with exactly three tracks:
a scattered positron
and two oppositely-charged hadrons.
The relevant 4-momenta are:
$k$ ($k'$) of the incident (scattered) positron,
$q \equiv k - k'$ of the virtual photon,
$P$ of the struck nucleon,
$P_{h^+}$ and $P_{h^-}$ of the detected hadrons,
$v \equiv P_{h^+} + P_{h^-}$ of the $\rho^0$ candidate,
and
$P_Y \equiv P + q - v$ of the undetected
final state $Y$.
The relevant Lorentz invariants are:
$Q^2 = -q^2 > 0$;
$\nu = q \cdot P/M$
(here $M$ is the proton mass);
an exclusivity measure
$\Delta E = (P_Y^2 - M^2)/2M$;
the invariant mass
$M_{\pi\pi} = \sqrt{v^2}$
assuming
the detected hadrons are pions;
the squared 4-momentum transfer $t = (q - v)^2$
to the target;
the maximum value $t_0$ of $t$ for fixed
$\nu$, $Q^2$, $P_Y^2$, and $M_{\pi\pi}$;
and the above-threshold momentum transfer $t' = t - t_0 < 0$.

For nuclear targets, the diffractive interaction with the target can occur
incoherently from individual nucleons
or coherently from the nucleus as a whole.
The incoherent exclusive $\rho^0$ production signal
was extracted
in the kinematic region
$t'_l < -t' < 0.4~\rm GeV^2$,
$-2{~\rm GeV} < \Delta E < 0.6~\rm GeV$,
$0.6{~\rm GeV} < M_{\pi\pi} < 1~\rm GeV$, and
$9{~\rm GeV}<\nu<20~\rm GeV$.
The lower $-t'$ limit, $t'_l$,
is chosen separately 
for each target and $l_c$ bin to maximize statistics
while keeping small the contribution from coherent scattering;
$t'_l$ is 0.03 to 0.06~GeV$^2$ for $^2$H,
0.03 to 0.14~GeV$^2$ for $^3$He, and
0.05 to 0.09~GeV$^2$ for $^{14}$N.

\begin{figure*}
\begin{center}
\includegraphics[width=0.47\textwidth]{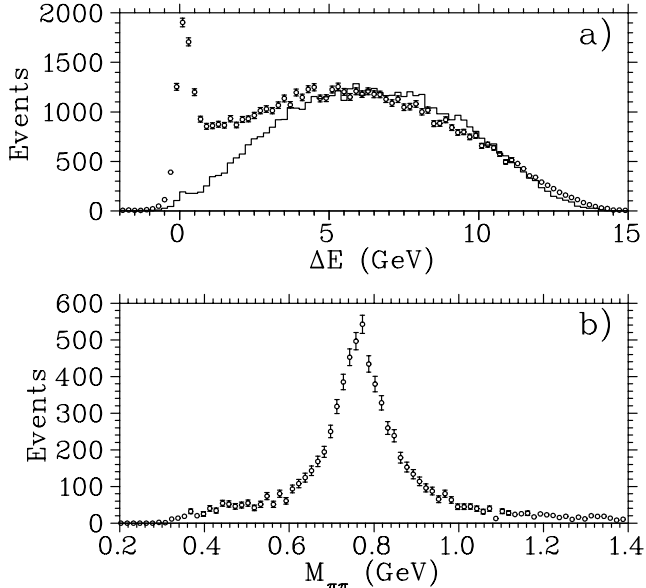}
\begin{minipage}[b]{0.48\textwidth}
\caption[]{\label{fig:de}\label{fig:m2pi}
(a) Measured events as a function of exclusivity variable $\Delta E$
for the 
$^1$H, $^2$H, $^3$He, and $^{14}$N data
passing the experimental
cuts; the distribution is shown for
$0.1{~\rm GeV^2} < -t' < 0.4~\rm GeV^2$ (open circles) and
for $0.7{~\rm GeV^2} < -t' < 5~\rm GeV^2$ (histogram, scaled
to the same total counts at $\Delta E > 3~\rm GeV$).
(b) Invariant mass distribution
for the exclusive
events at $0.1{~\rm GeV^2}<-t'<0.4~\rm GeV^2$.
}
\end{minipage}
\end{center}
\end{figure*}

The distribution of the selected events in $\Delta E$
is shown for all targets in
Figure~\ref{fig:de}a.
Exclusive
$eN\rightarrow eh^+h^-N$
events, where
the undetected final state
consists of a nucleon recoiling without excitation,
occur at $\Delta E = 0$.
{\em Non-exclusive}
events that involve the production
of additional, undetected particles
appear at larger $\Delta E$.
The events with
$\Delta E \gtrsim 3~\rm GeV$ are
predominantly due to deep inelastic scattering (DIS).
The $\Delta E$ dependence of DIS events is measured
at
$0.7 < -t' < 5~\rm GeV^2$
where the diffractive exclusive signal is negligible
(see histogram in
Figure~\ref{fig:de}a).
The DIS background below the exclusive peak
is subtracted for each target and
kinematic bin separately,
assuming the shape of the background is independent
of $t'$
and normalizing to
the number of events measured
at
$t'_l < -t' < 0.4~\rm GeV^2$ and
$\Delta E>3~\rm GeV$.
The difference 
at $\Delta E \sim 2~\rm GeV$
between the two distributions
shown in Figure~\ref{fig:de}a is due mainly to
the radiative tail of the exclusive peak and to
$\rho^0$ production events where the diffractive interaction
excites the nucleon.  Except for small kinematic shifts,
these processes do not affect the propagation of the
virtual photon or outgoing $\rho^0$ through the nuclear medium.

The exclusive $M_{\pi\pi}$ distribution,
shown in
Figure~\ref{fig:m2pi}b,
is dominated by resonant production of the $\rho^0(770)$,
with small interfering contributions from
exclusive production of
non-resonant $\pi^+\pi^-$ pairs and
of the $\omega(782)$ resonance
(in its 2\% decay branch
to $\pi^+\pi^-$\cite{pdb}).
Background
from the two-kaon decay of exclusively-produced $\phi(1020)$ mesons,
which
would appear at
$M_{\pi\pi}<0.5~\rm GeV$,
is eliminated
by requiring that the two-kaon invariant mass be greater
than
$1.04~\rm GeV$.

\begin{figure*}
\begin{center}
\includegraphics[width=0.47\textwidth]{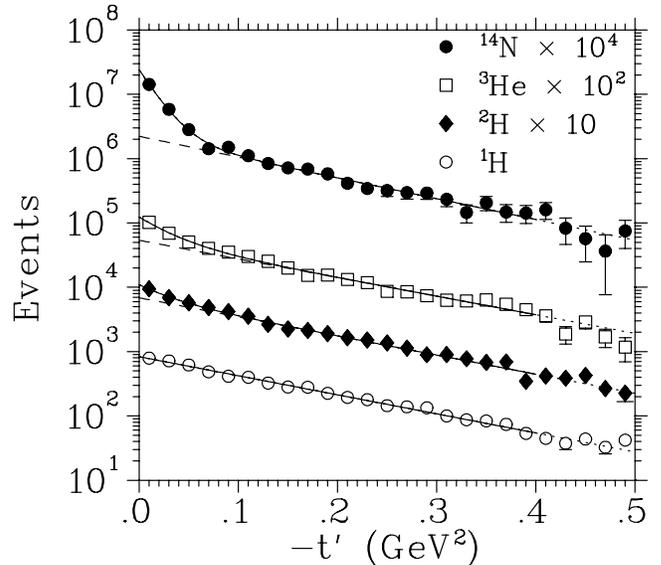}
\begin{minipage}[b]{0.48\textwidth}
\caption[]{\label{fig:t1}
Distribution of momentum transfer $t'$
for exclusive $\rho^0$-production
from
$^1$H, $^2$H, $^3$He, and $^{14}$N
targets.
The solid curves are fit to
$a_N (b_N e^{b_N t'} + f_A b_A e^{b_A t'})$,
the dotted lines are extrapolations beyond the fit
interval $-t'<0.4~\rm GeV^2$,
and the dashed lines are the
inferred incoherent
contributions.
}
\end{minipage}
\end{center}
\end{figure*}

The exclusive $-t'$ distributions
for the $^1$H, $^2$H, $^3$He, and $^{14}$N nuclei
are shown in
Figure~\ref{fig:t1}.
The data exhibit the rapid falloff expected for
a diffractive process.
To isolate incoherent
scattering, the data are fit to a shape giving the sum of
incoherent and coherent contributions,
$b_N e^{b_N t'} + f_A b_A e^{b_A t'}$ (solid curves).
Here $f_A$ is the ratio of coherent to incoherent total counts
and $e^{b_Nt'}$ ($e^{b_At'}$) represents the product of the
$\rho^0$ and struck nucleon (nucleus)
elastic form factors, squared\cite{povh}.
The incoherent slope parameter $b_N$ for each nucleus
(measured to an accuracy of about $0.5~\rm GeV^{-2}$)
is
consistent
with the hydrogen value
$b_N = (6.82 \pm 0.15)~\rm GeV^{-2}$.
The coherent slope parameters 
$b_{^2\rm H} = (33.3 \pm 9.8)~\rm GeV^{-2}$,
$b_{^3\rm He} = (32.5\pm 5.7)~\rm GeV^{-2}$, and
$b_{^{14}\rm N} = (57.2\pm 3.3)~\rm GeV^{-2}$
are consistent with the values
predicted by the
relationship
$b_A \approx R_A^2/3$\cite{povh}
and
the measured electromagnetic RMS radii
$R_{^2\rm H} = 2.1~\rm fm$,
$R_{^3\rm He} = 1.9~\rm fm$, and
$R_{^{14}\rm N} = 2.5~\rm fm$\cite{rms}.

In the absence of
ISI and FSI,
the 
cross section
$\sigma_A$ for
incoherent $\rho^0$ production
from a nucleus with $A$ nucleons
would be
$A \sigma_H$
(assuming the expected isospin symmetry
$\sigma_n = \sigma_H$\cite{bauer}, where $n$ and $H$ refer
to the neutron and $^1$H).
The
nuclear transparency is therefore
$T_A \equiv \sigma_A/(A \sigma_H) = N_A L_H/(A N_H L_A)$,
where the second equality follows from the $A$-independence of
the experimental acceptance.
Here $N_{A,H}$ is the number of incoherent events in the range
$t'_l < -t' < 0.4~\rm GeV^2$;
$N_A$ is corrected for the coherent contribution using the $t'$ fit
for each $l_c$ bin
($t'_l$ is chosen so
that the correction factor
is less than 1.05
with an uncertainty of less than 4\%).
The integral
$L_{A,H}$ of the effective
luminosity
is determined from the number of inclusive DIS positrons
and the published nuclear DIS structure functions\cite{sigdis},
with a correction 
for the efficiency
($\gtrsim 0.8$)
for tracking the $h^+h^-$ pair.

The dominant systematic uncertainties are from possible
differences
in the spectrometer performance for the nuclear and $^1$H
data (estimated by studying the time dependence of $N_{A,H}/L_{A,H}$ and
other normalized yields)
and from
the treatment of the non-exclusive
background (estimated by studying the dependence of $T_A$ on $\Delta E$).
The systematic uncertainty
in the overall normalization of
$T_{^2\rm H}$, $T_{^3\rm He}$, or $T_{^{14}\rm N}$
is 2.7\%, 5.5\%, or 5.9\% respectively.
The additional point-to-point systematic uncertainty includes
the fit uncertainty in the coherent contribution.
The $T_A$ results are unchanged
at the 3\% level (and the systematic uncertainties
are essentially unchanged)
if the
non-exclusive background is not subtracted.

\begin{figure*}
\begin{center}
\includegraphics[width=0.47\textwidth]{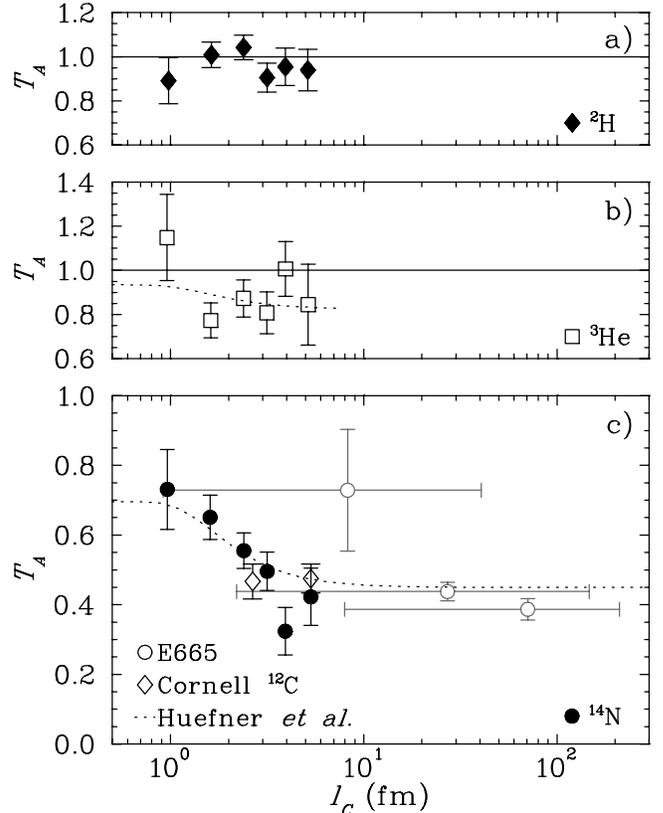}
\begin{minipage}[b]{0.48\textwidth}
\caption[]{\label{fig:ta-lcoh}
Nuclear transparency $T_A$ as a function of $l_c$ for
a) $^2$H (filled diamond),
b) $^3$He (open square),
and c) $^{14}$N (filled circle) targets.
The error bars include
statistical and point-to-point systematic uncertainties added in quadrature.
The systematic
uncertainty in
the overall normalization of $T_A$ is not shown.
Panel (c) includes comparisons with previous expriments with
photon
(open diamonds)\cite{mcclellan}
and muon (open circle)\cite{e665}
beams.
Due to the acceptance for $20<\nu\lesssim 370~\rm GeV$,
the three $Q^2$ bins measured by\cite{e665}
correspond to broad ranges in $l_c$
(horizontal error bars).
The dashed curves are the
Glauber
calculation of
H\"ufner {\it et al.} for $^3$He and $^{14}$N\cite{kopel}.
}
\end{minipage}
\end{center}
\end{figure*}

The nuclear transparencies
for $^2$H (filled diamond), $^3$He (open square),
and $^{14}$N (filled circle)
are shown
as
functions of the coherence length $l_c$
in
Figure~\ref{fig:ta-lcoh}.
Within uncertainties the $^2$H and $^3$He transparencies are
independent of $l_c$:
$T_{^2\rm H} = 0.970\pm 0.024$ (statistical) $\pm 0.040$ (systematic)
and
$T_{^3\rm He} = 0.862\pm 0.042\pm 0.061$.
The consistency of the deuterium transparency with unity
suggests that $\sigma_n \approx \sigma_H$
and that
the ISI and FSI are small in $^2$H.
The average $^3$He transparency is
1.9 standard deviations below unity.

The 
nitrogen transparency exhibits the decrease expected
from the onset of hadronic ISI
as $l_c$ increases.
The decrease from
$0.681 \pm 0.060$ at
$l_c<2~\rm fm$ to
$0.401 \pm 0.054$
at
$l_c>3.6~\rm fm$
(errors exclude normalization uncertainty)
has a 3.5 standard deviation
statistical significance.
In the absence of ISI variations,
the transparency would exhibit a
small ($< 3\%$)
{\em increase} with $l_c$
due to
the known\cite{bauer}
energy dependence of the $\rho^0N$ cross section.

Figure~\ref{fig:ta-lcoh}c
also shows
the transparency to incoherent $\rho^0$ production
measured at Cornell with 4 and 8~GeV photons\cite{mcclellan}
and by the E665 collaboration at FNAL with 470~GeV muons\cite{e665}.
These results
are consistent with the present data
but give no indication of a variation with $l_c$.
The E665 $T_{^{14}\rm N}$ values
are inferred from the
published $A$-dependence\cite{e665}.
The E665
value for
$T_{^{14}\rm N}$
at $l_c\sim 8~\rm fm$
was measured
at $\nu\gtrsim 100~\rm GeV$ and $Q^2>3~\rm GeV^2$\cite{e665}, and
may therefore be influenced by
color transparency.
Color transparency
implies
that at high $Q^2$ and $\nu$
the $q\bar{q}$ pair (and the subsequent $\rho^0$)
is produced and propagates in a non-interacting
configuration of reduced transverse size,
resulting in
$T_A \rightarrow 1$\cite{bauer,qcdmodel,ct,kopelct}.
For this reason
data collected by the NMC collaboration with a muon beam
at $40{~\rm GeV}<\nu<180~\rm GeV$ and $Q^2>2~\rm GeV^2$\cite{nmc}
are not included in
Figure~\ref{fig:ta-lcoh}c.

The 
$T_{^{14}\rm N}$ and
$T_{^3\rm He}$
data are consistent with a recent prediction
(dashed curves in Figure~\ref{fig:ta-lcoh})
of the coherence
length effect\cite{kopel},
although the statistics for
$T_{^3\rm He}$ are not sufficient to demonstrate the $l_c$ variation.
The
prediction uses Glauber multiple-scattering theory\cite{glauber},
where the total $\rho^0$ production amplitude is the sum
of the amplitudes from each nucleon, modified by
elastic and
inelastic rescattering
of the outgoing $\rho^0$ on the other nucleons.
In this model, the
$q\bar{q}$ fluctuation from which the $\rho^0$ originates
is found to
interact with the nuclear medium like a $\rho^0$\cite{kopel}.
The strength of the $\rho^0$ and $q\bar{q}$
interactions govern the transparency
at small $l_c$ and its $l_c$ dependence, respectively.
The consistency of the model with the data therefore suggests
that when $l_c$ is large, the $q\bar{q}$ ISI are
approximately as strong as the $\rho^0$ FSI.
For the $\nu$ values of the present measurement,
color transparency is expected to produce
little deviation from the Glauber prediction
\cite{kopel,kopelct}.

The data support the
hypothesis\cite{bauer,badelek}
that absorption of the photon's $q\bar{q}$ component
contributes to the shadowing observed in
real and virtual photon nuclear cross sections.
Shadowing denotes that the
cross sections grow
more slowly than linearly in $A$.
It
is observed
for
inclusive DIS at small
Bjorken $x = Q^2/2 M \nu$
and for elastic and
inclusive real photon scattering at high energies.

In summary, the transparency of the $^2$H, $^3$He, and $^{14}$N
nuclei to
exclusive incoherent $\rho^0$ electroproduction
was measured by the HERMES experiment
as a function of the coherence length
of $q\bar{q}$ fluctuations of the virtual photon.
The measured transparencies agree well with
previous data and with
a prediction using the
standard treatment of high-energy initial and final state interactions.
The transparency of the
nitrogen nucleus exhibits a significant
decrease with $l_c$, which
is attributed
to initial state interactions of the
$q\bar{q}$ fluctuation from which
the $\rho^0$ originates.

We gratefully acknowledge the DESY management for its support and
the DESY staff and the staffs of the collaborating institutions.
We would also like to thank B.Z. Kopeliovich,
J. Nemchik, M. Strikman, and especially D.F. Geesaman for
helpful discussions.
This work was supported by
the FWO-Flanders, Belgium;
the Natural Sciences and Engineering Research Council of Canada;
the INTAS and TMR network contributions from the European Community;
the German Bundesministerium f\"ur Bildung, Wissenschaft, Forschung
und Technologie; the Deutscher Akademischer Austauschdienst (DAAD);
the Italian Istituto Nazionale di Fisica Nucleare (INFN);
Monbusho, JSPS, and Toray
Science Foundation of Japan;
the Dutch Foundation for Fundamenteel Onderzoek der Materie (FOM);
the U.K. Particle Physics and Astronomy Research Council; and
the U.S. Department of Energy and National Science Foundation.

\end{multicols}

\end{document}